# Management of mobile resources in Physical Internet logistic models

Jean-Yves Colin, Moustafa Nakechbandi
LITIS, Le Havre University, 5 rue Ph. Lebon,
Le Havre, France
{moustafa.nakechbandi, jean-yves.colin}@univ-lehavre.fr

Hervé Mathieu
ISEL, Quai Frissard
Le Havre, France
herve.mathieu@univ-lehavre.fr

***Abstract*— This paper deals with the concept of a "Physical Internet", the idea of building large logistics systems like the very successful Digital Internet network. The idea is to handle mobile resources, such as containers, just like Internet data packets. Thus, it is possible to use the principles of encapsulation and routing to optimize the freight. The problem is that mobile resources, such as containers, are not quite similar to data packets, because they are real and not dematerialized. Thus the handling and the storing of mobile resources, such as containers, will create imbalances in the logistics network, leading to starvation or overstocking of logistic network nodes. We propose in this paper a study addressing this problem leading to some solutions.**

## I. Introduction

Although logistical systems are now almost everywhere and are essential to modern societies, they are full of inefficiencies from the economic, environmental and societal points of view [8]. New ideas and paradigms, sometimes borrowed or inspired by other fields, may help to improve them. Among them is the concept of a "Physical Internet", the idea of building large logistics systems like the very successful Digital Internet network.

A Physical Internet (PI) [7] is an open global logistic system founded on physical, digital and operational interconnectivity through encapsulation, interfaces and protocols, such as data packets in the Digital Internet. The term Physical Internet employs a metaphor taken from the Digital Internet, which is based on routers, all transmitting standard packets of data under the TCP-IP protocol. A core enabling technology to make the PI a real "tour de force" is the encapsulation of goods in modular, reusable and smart containers, the π-containers (physical freight containers). The π-containers range in modular dimensions from small ones to large ones. The ubiquitous usage of π-containers will make it possible for any company to handle and store any products because they will not be handling and storing products per se. Instead they will be handling standardized modular containers, just as the Digital Internet transmits data packets rather than information/files. In the rest of this paper, we are interested in finding an efficient method to handle mobile resources where there is an imbalance to be straightened.

## II. Problem Description

We suppose that we deal with standard containers of different size which can be handled like conventional IP packets [3]. The problem is that we cannot handle physical containers (which are expensive and taking lots of space) just like IP packets. Indeed, IP packets can be thrown away and created again from nothing, which is not the case of π-containers.

Thus, an important problem is that unbalanced logistics flows will induce the gathering of those mobile resources at specific locations, whereas they would be needed in some other places. If it is possible to easily create a data packet from nothing containing data to be transmitted, the lack of appropriate π-containers at one node will prevent expeditions on the PI from that node. Thus, there can be too many mobile resources at one node and not enough at one other.

Mobile resources management in classical logistics networks is the responsibility of the owner of those resources. Management is barely mutualized. It is necessary in the PI concept to globally manage these mobile resources, and not to consider them from an individual point of view.

## III. The Model

In the model we consider a graph $(V,E)$ representing the studied area. $V$ is a set of locations (ports, hub...), representing producing units (factories...), and consuming units (cities, factories...), $E$ is the set of physical paths between the locations in $V$. $\{d_{i,j} ; (i,j) \in E\}$ represents the distance between $i$ and $j$.

To make it simpler, we will consider only one kind of container, for example, the refrigerated "reefers" for food transportation. Each location $v$ in $V$ may receive, store and send these containers. Each location starts with some number $\mathcal{X}_v$ of these containers, and we call $N$ the total number of reefers. Due to the usual imbalance between the needs and the offers at each location, it is normal for the containers to be accumulated at the mostly consuming locations, and to disappear from producing ones. At consuming locations, storing unused reefers is costly, so a policy of sending away unused reefers is necessary.

Also, factories will have to ask to get some new reefers or bring back empty ones to send their products to consuming locations. We assume the presence of an efficient policy to decide how to manage this imbalance. According to this policy, a number $y_v$ of reefers is necessary at each location *v*. This state is supposed to be reachable. We have also $\sum_{v \in V} x_v = \sum_{v \in V} y_v$ .

An efficient way to direct the reefers must then be used. We have one ship, with its starting location, and its limited capacity *k*. Now we must decide how to use this ship and set the path it should follow so that, at the end, the planed number of reefers $y_v$ is available at each location *v*.

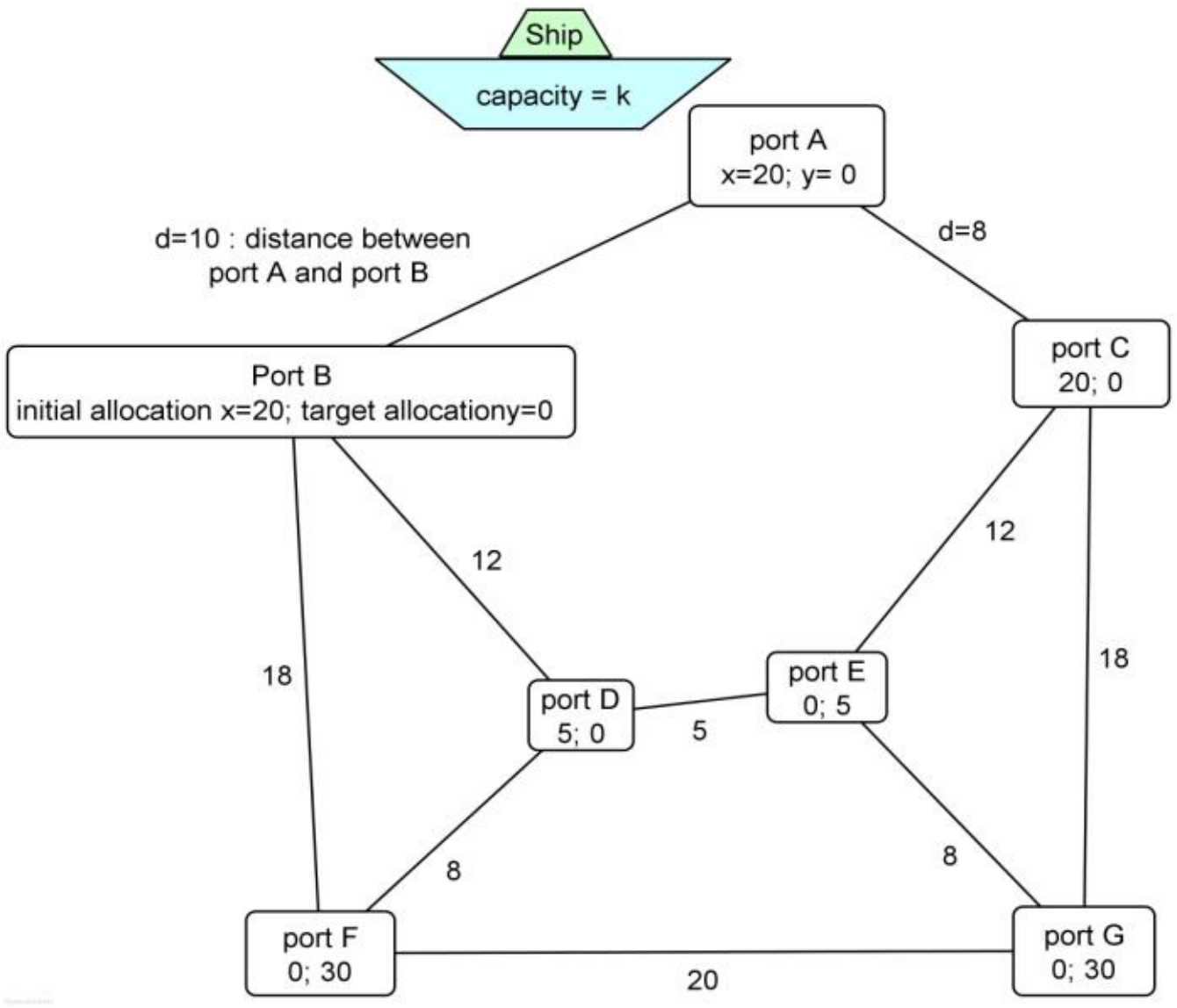

Figure 1. Example of graph in which the vertices are locations, and edges are paths. Each vertex has a value x of current resources, and a value y of needed resources. Each edge is valued by its length d.

So our goal is to find a sequence *S* of visited stations (ports) and the resources displacements that brings the system from the initial state *X* to the desired state *Y* such that the total traveled distance in sequence S is as small as possible.

We suppose that the shortest paths between all pairs of vertices is known (using Roy–Warshall algorithm [9], for example), and that, for each location, the paths from this node to all the other locations are sorted by increasing lengths.

## IV. Solution

There are many similar problems already studied : Bipartite TSP, Delivery Problem, Swapping Problem that have been the subject of extensive studies in the literature [1, 2, 5]. All these problems are known to be NP-hard problems, and so is this problem. Because the above problem is a NP-hard problem, we propose to use the heuristic presented in [4] that is in *O(log k + n logN)* where *k* is the capacity of the ship, *n* is the number of stations, and *N* is the total number of mobile resources to be moved.

This heuristic can be divided in the following four steps :

- The first step computes a matching *M* of minimal costs between the vertices that have too many resources and need less or none, and vertices that have too few resources and need more of them.
- The second step builds two tours : The first one a tour $C^{ex}$ that passes through all the vertices that have too many resources and need less or none. The second tour $C^{def}$ that passes through all the vertices that have too few resources and need more of them.
- The third step splits the $C^{ex}$ and $C^{def}$ tours into subpaths of multiple of the capacity *k* of the ship, some of these subpaths having too many resources, and the others needing resources.
- The fourth and last step transfers resources through the b-matching *M* from the subpaths that have too many resources, to the other subpaths that need resources.

Each step must take into account the lengths between vertices, so it uses a nearest neighbor heuristic to try to keep the total length small when building the intermediate results.

We will use the example of Figure 1 to illustrate the computation of this heuristic. In this example we suppose *k=5* and we see that the availability and needs are multiples of the k capacity. The following table gives the distances from each vertex to each other vertex. It will be used by nearest neighbor heuristic when choosing among several possible destinations.

Table 1: Length of shortest paths between all pairs of vertices in Figure 1.

| | **A** | **B** | **C** | **D** | **E** | **F** | **G** |
|---|---|---|---|---|---|---|---|
| **A** | 0 | 10 | 8 | 22 | 20 | 28 | 26 |
| **B** | 10 | 0 | 18 | 12 | 17 | 18 | 25 |
| **C** | 8 | 18 | 0 | 17 | 12 | 25 | 18 |
| **D** | 22 | 12 | 17 | 0 | 5 | 8 | 12 |
| **E** | 20 | 17 | 12 | 5 | 0 | 12 | 8 |
| **F** | 28 | 18 | 25 | 8 | 12 | 0 | 20 |
| **G** | 26 | 25 | 18 | 12 | 8 | 20 | 0 |

The first step computes a perfect b-matching between excess vertices and default vertices, using a nearest neighbor heuristic to choose when several solutions are possible (figure 2).

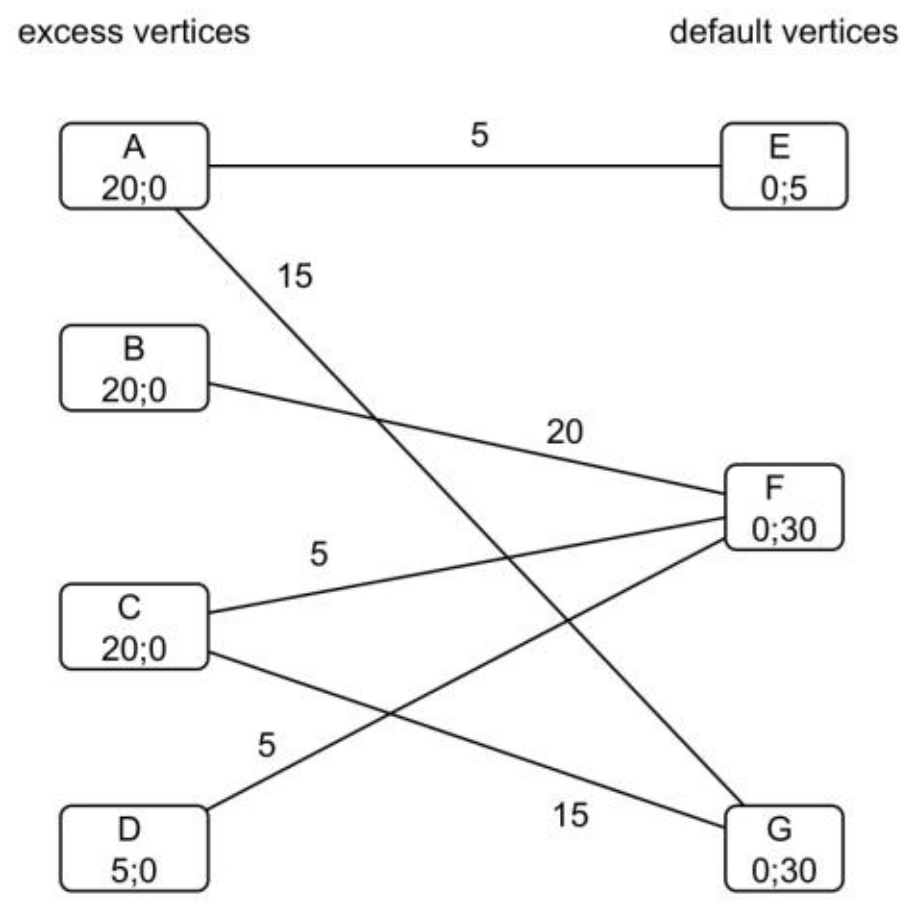

Figure 2. b-matching M between excess vertices and default vertices for the input of Figure 1

Step 2 builds a tour $C^{ex}$ that passes through all the vertices that have too many resources and need less or none. It then builds a tour $C^{def}$ that passes through all the vertices that need resources and have a few or none (again, a nearest neighbor heuristic is used to choose between several possibilities when building each tour) (figure 3).

Note that in our example, there is no need to divide the vertices because each one has a value that is a multiple of *k*. When this is not the case, these vertices must be split to be included into paths in the tour, so that each path has a total value that is a multiple of *k*.

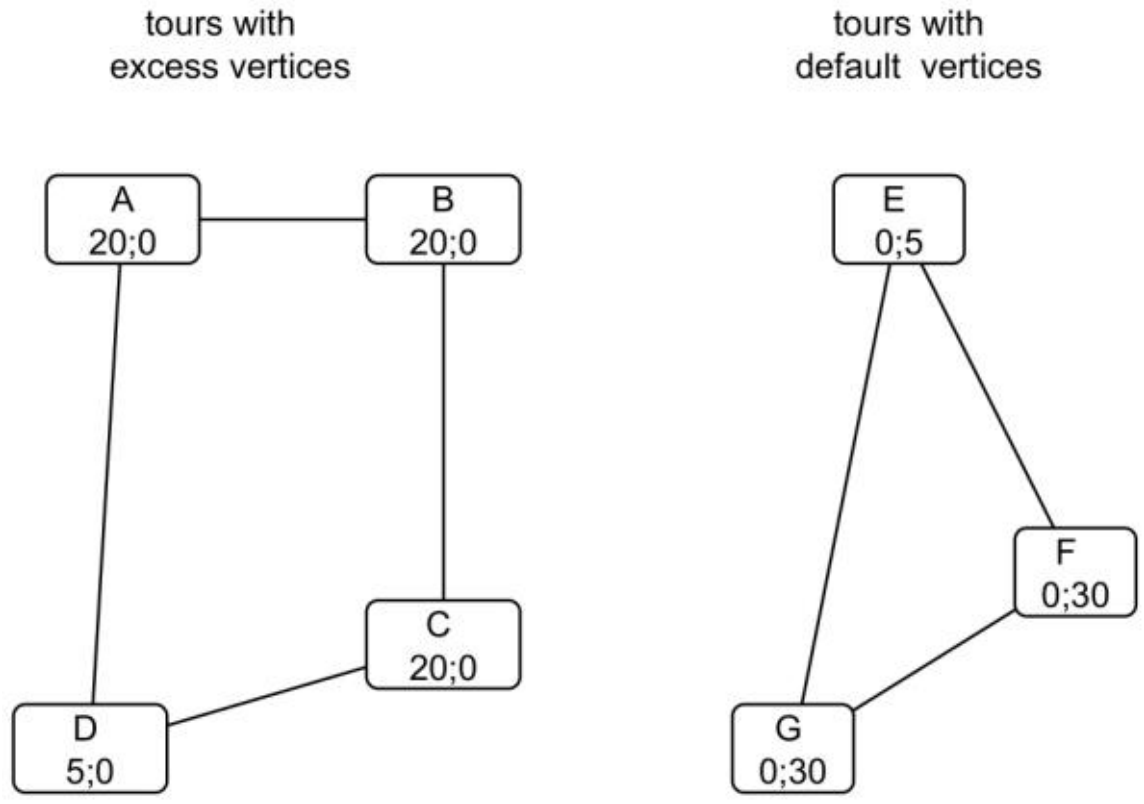

Figure 3. Builds two tours : the firest passing through all excess vertices, the second passing through all default vertices.

Step 3 builds a super-matching graph labeled by multiple of *k* containers (*k=5*) compatible with the step 1(figure 2) and the step 2 (figure 3). The result is presented in figure 4.

Note that if the value of a vertex was not a multiple of *k*, this vertex of the original graph in figure 1 would be split into several vertices in figure 2 and would appear in several vertices of figure 3. The super maching graph in figure 4 would then be different from the b-maching graph in figure 2.

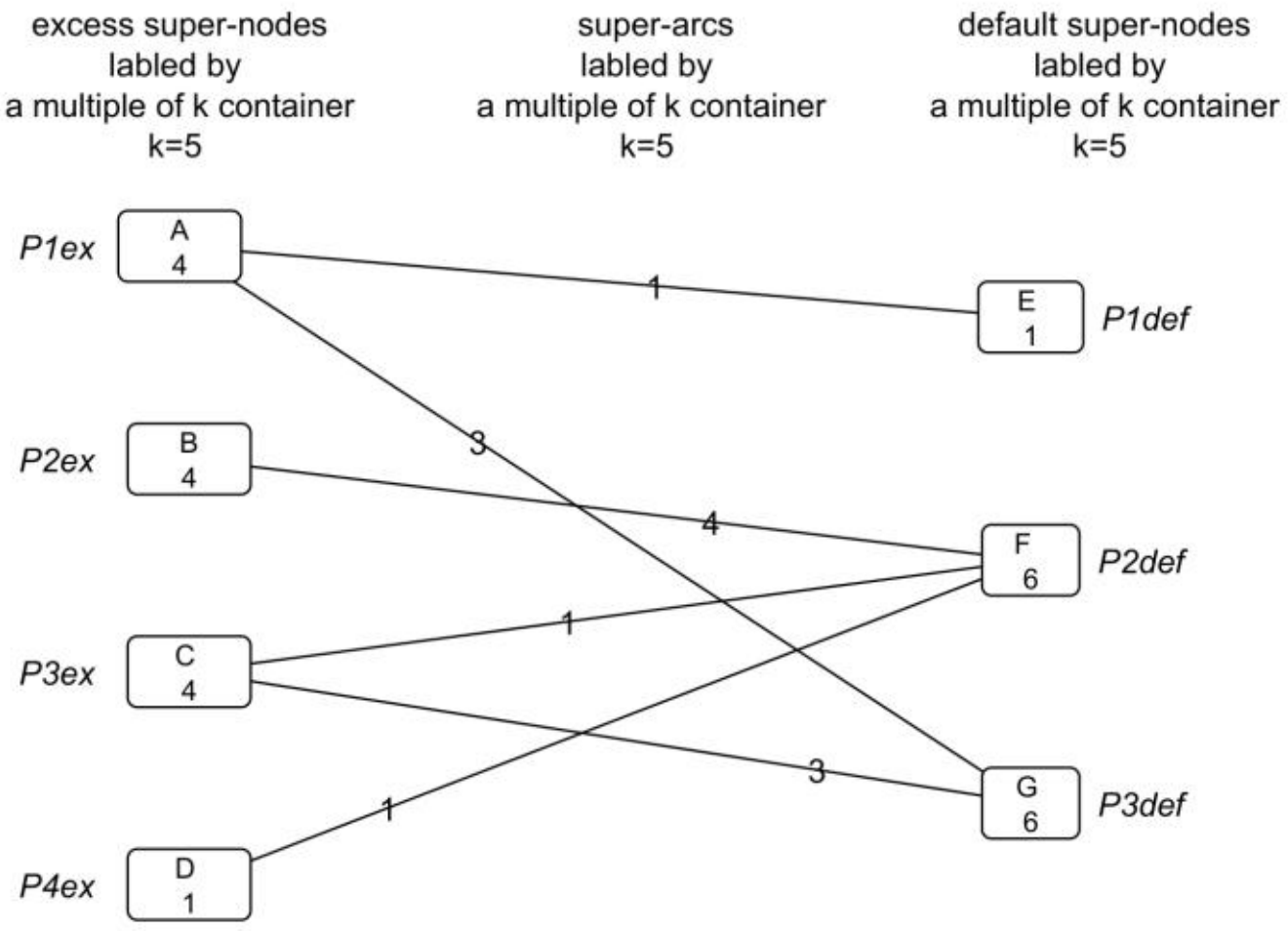

Figure 4. Construction of the super-matching graph for *k=5*. Each super-node here represents a group of one or more vertices in Figure 1.

Step 4 Everything is put together to transfer containers from vertices with too many containers (excess in containers) to vertices with too few of them (default in containers). Induced by the super-matching of figure 4, and starting arbitrarily from vertex A in $P_1^{ex}$ (noted *P1ex* in the figure 4), a path is build that first ships *k* containers to E in $P_1^{def}$ (because E is closer from A than is G, the other vertex associated to A in the super-matching of Figure 4) once with containers from A in $P_1^{ex}$. Then it ships *k* containers to G in $P_3^{def}$ thrice with containers left from $P_1^{ex}$ (because no other vertex needing containers is associated to a in the super matching). Then it goes to $P_3^{ex}$. (because C is the only vertex with excess containers associated to G). So it ships *k* containers to G in $P_3^{ex}$ from C. etc.

The final path that is a solution to the proposed problem is the following : we start from A. The first step is to go to E from A. We go back to A, and then, we perform 3 round trips between A and G, until A is empty. We go then from G to C. There are 3 round trips between C and G. The journey (not empty) will be from C to F, then one round trip from F to D. At the end, 4 round trips from B to F, and we finish at F. In summary : (A, E, A, G, A, G, A, G, C, G, C, G, C, F, D, B, F , B, F, B, F, B, F). It is represented in figure 5.

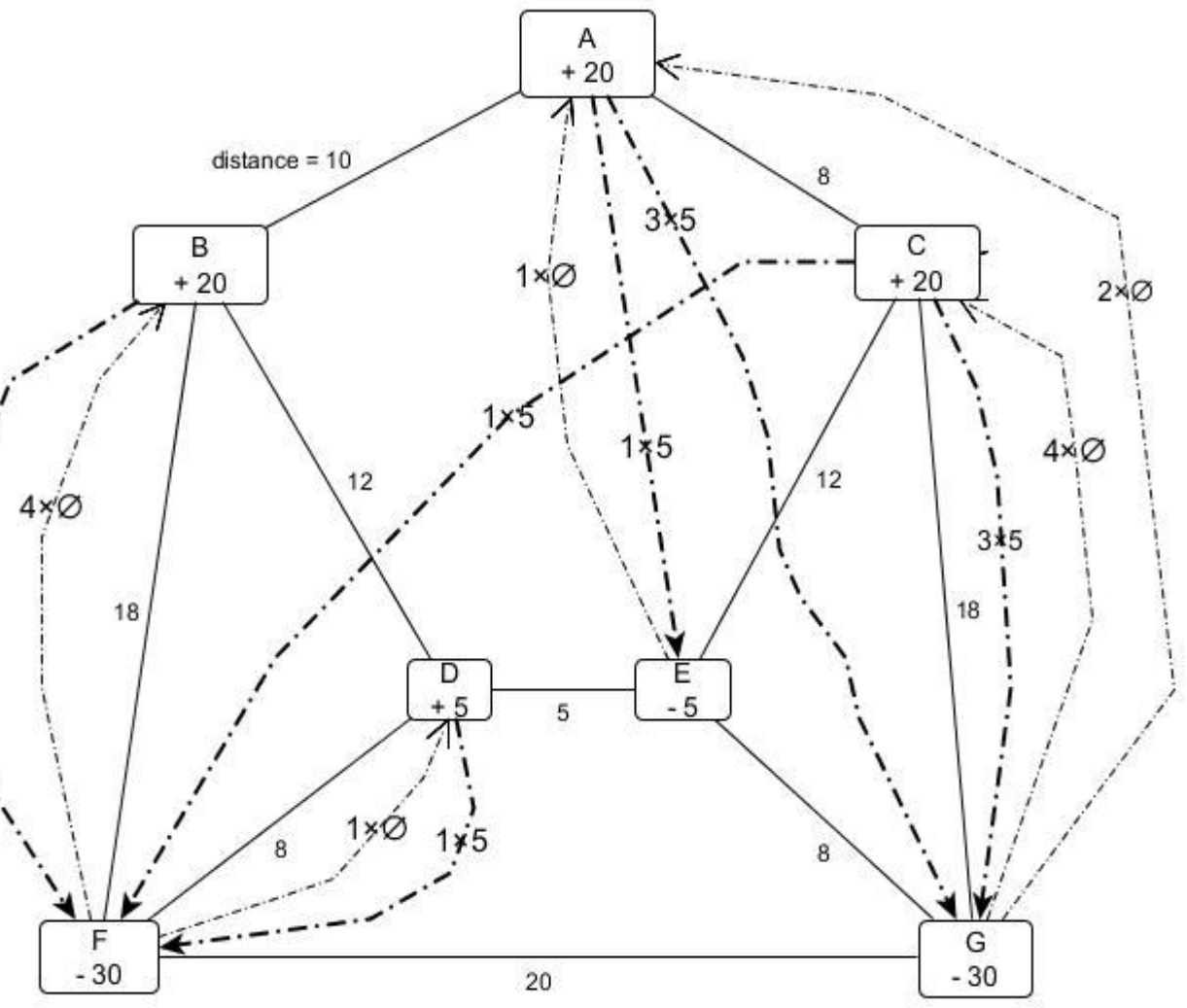

Figure 5. A solution induced by the super-matching of Figure 4. Labels in vertices are exess in containers if they are positive, and are default in containers if they are negative. A thick dashed arc labeled by $p\times k$ represents p shippings of *k* containers. A thin dotted arc labeled by $q\times\varnothing$ represents q return trips without any containers.

## V. Conclusion and Remarks

In this paper, we exposed the problem of efficiently handling mobile resources, such as containers, based on the Physical Internet concept. Indeed, the handling and the storing of mobile resources (containers...) will create imbalances in the logistic network, leading to starvation or overstocking in logistic network nodes. After presenting a model of mobile resource imbalance, we explored some ideas to straighten imbalances during the mobile resources management. In the future we intend to study the interactions between several areas, each one with its own graph, and the problem in which there are several ships available to handle the mobile resources.

## ACKNOWLEDGEMENT

This work is supported by the Haute-Normandie Region : Projet CLASSE "Corridors Logistiques : Applications à la vallée de la Seine et Son Environnement", France.